\documentclass[pra,aps,showpacs,twocolumn]{revtex4}
\usepackage{epsfig}

\begin{document}
%\setstretch{2} 

\title{Naturally-phasematched second harmonic generation in a whispering gallery mode resonator}

\author{J. U. F{\"u}rst$^{1,2}$}
\author{D. V. Strekalov$^{1,3}$}
\author{D. Elser$^{1,2}$}
\author{M. Lassen$^{1,4}$}
\author{U. L. Andersen$^{1,4}$}
\author{C. Marquardt$^{1,2}$}
\author{G. Leuchs$^{1,2}$}

\affiliation{$^1$Max Planck Institute for the Science of Light, Erlangen, Germany\\ $^2$Department of Physics, University of Erlangen-Nuremberg, Germany\\
$^3$Jet Propulsion Laboratory, California Institute of
Technology, Pasadena, California, USA\\
$^4$Department of Physics, Technical University of Denmark, Kongens Lyngby, Denmark}

\date{\today}

\begin{abstract}
We demonstrate for the first time natural phase matching for optical frequency doubling in a high-$Q$ whispering gallery mode resonator made of Lithium Niobate. A conversion efficiency of 9\% is achieved at $30\;\mu$W in-coupled continuous wave pump power. The observed saturation pump power of 3.2 mW is almost two orders of magnitude lower than the state-of-the-art. This suggests an application of our frequency doubler as a source of non-classical light requiring only a low-power pump, which easily can be quantum noise limited.  Our theoretical analysis of the three-wave mixing in a whispering gallery mode resonator provides the relative conversion efficiencies for frequency doubling in various modes. 
%\noindent$OCIS: 190.2620,\;230.5750,\; 190.4360,\;230.4320 $
\end{abstract}

\maketitle

Various types of oscillations can be enhanced by appropriate resonators. Examples range from resonating chambers for sound waves to cavities
for electromagnetic waves. Nonlinear cross-coupling of different oscillations can be enhanced as well, e.g. between optical waves, between optical and radio-frequency electromagnetic waves, and, a subject of considerable recent attention, between mechanical and optical oscillations \cite{optomech}. 

A resonator of particular interest in the context of quantum and nonlinear optics is the whispering gallery mode resonator (WGMR). In WGMRs, light is guided via continuous total internal reflection. The small mode volume combined with the absence of external reflecting elements provides for a very high quality factor $Q$ and good spatial localization of the optical fields. 
In addition to these important capabilities, the WGMR enables us to continuously control the coupling strength between the intra-cavity modes and the modes outside the cavity. This capability may be compared to a cavity mirror with continously variable reflectivity. 

Due to these unique capabilities, various nonlinear processes have been observed in WGMRs, including four wave
mixing \cite{savchenkov04,kippenberg04,delhaye07comb,savchenkov08comb}, Raman \cite{spillane02,savchenkov08comb}, parametric \cite{savchenkov07} and Brillouin \cite{grudinin09} scattering, microwave upconversion \cite{Strekalov09THz}, second \cite{ilchenko04SH} and third \cite{carmon07} harmonic generation.
Furthermore, a plethora of dynamical effects has been predicted in the cavities with the second order nonlinear response, examples being bistability, chaotic behavior
and self pulsing \cite{Drummond80Marte94}, along with non classical effects such as squeezing and entanglement \cite{Drummond81}. It is intriguing to investigate these effects in a highly efficient WGMR. 

In this paper 
we report on the second harmonic (SH) generation in a doubly resonant  WGMR. We employ non-critical type-I phase matching, and achieve the SH conversion efficiency of $9\%$ at 30 $\mu$W pump power. This is nearly two orders of magnitude higher than the previous state-of-the-art result for the same pump power, obtained with a less efficient quasi phase matching \cite{ilchenko04SH}. At higher pump power, the conversion efficiency saturates due to a dynamic ``self-limiting" effect. 

The phase matching conditions for SH and degenerate parametric
down conversion (PDC) reflect the conservation laws associated with the resonator symmetries. They arise from the interaction Hamiltonian
\begin{equation}
H_{int}=\int\chi^{(2)}E_{s}(E^\dagger_p)^2dV+h.c.
\label{H}
\end{equation} 
expressed via the effective quadratic nonlinearity $\chi^{(2)}$ and the pump and SH electric fields $E_p$ and $E_{s}$. In Cartesian coordinates, the integral in (\ref{H}) gives rise to three momentum conservation equations: $2{\vec k}_p={\vec k}_{s}$. In a spherical WGMR, a mode is defined by a set of polar, azimuthal and radial numbers $\{L,m,q\}$. Assuming that the WGMR radius $R$ is much larger than the optical wavelength $\lambda$, such that there is no coupling disturbance, and that the WGMs are near-equatorial, we can reduce the electric fields in both polarizations to a scalar eigenfunction \cite{Kozyreff08} with a scaling factor $E_0:$
\begin{equation}
E_{Lmq}(r,\theta,\phi)= E_{0} Y_{Lm}(\theta,\phi)
j_{L}(k_{Lq}r).
\label{E}
\end{equation}
The angular part of (\ref{E}) is a spherical harmonic
\begin{equation}
Y_{Lm}(\theta,\phi)=
\sqrt{\frac{(2L+1)(L-m)!}{4\pi(L+m)!}}
P^m_{L}(\cos\theta)e^{im\phi},
\label{Esph}
\end{equation}
where $P^m_{L}(\cos\theta)$ are the associated Legendre polynomials.
The radial wave number $k_{Lq}$ in the argument of the spherical Bessel function $j_{L}(k_{Lq}r)$ relates the WGM eigenfrequency $\omega$ to $L$ and $q$ via the dispersion equation
\begin{equation}
k_{Lq}= \omega\frac{n}{c} = \frac{L}{R}\left[1+\alpha_q(2L^2)^{-1/3}+O(L^{-1})\right],
\label{disp}
\end{equation}
where $n$ is the material index of refraction
and $\alpha_q$ is the $q$-th zero of the Airy function \cite{Kozyreff08,ilchenko03dispersion}.

Applying (\ref{H}) to the angular part of (\ref{E}) yields the Clebsch-Gordan coefficients whose subscripts $s$ and $p$ refer to the pump and the second harmonic, respectively:
\begin{eqnarray}
\langle L_p ,L_p; m_p ,m_p&|& L_{s},m_{s}\rangle = \label{sigmaL}\\ 
= &&\int%_0^\pi\int_0^{2\pi}
Y^*_{L_{s}m_{s}}(\theta,\phi)Y^2_{L_pm_p}(\theta,\phi)\sin\theta d\theta d\phi. \nonumber
\end{eqnarray}
Selection rules for Clebsch-Gordan coefficients are well-known. In our case:
\begin{equation}
m_{s}=2m_p, \quad L_{s}\leq 2L_p. \label{m}
\end{equation}
The first condition in (\ref{m}) is conservation of the azimuthal component of the orbital momentum, while the second one arises from the triangular condition for a vector sum of the orbital momenta of two pump photons. The phase matching conditions for the SH generated on a \emph{surface} of a spherical WGMR have been already construed as the addition rules for the orbital momenta \cite{Kozyreff08}. This also applies to our WGMR, and the Clebsch-Gordan factors (\ref{sigmaL}) in the conversion efficiency emphasize the analogy with atomic systems.   
Notice that a spectrum of a spherical WGMR (\ref{disp}) is degenerate with respect to $m$, which decouples the first selection rule of (\ref{m}) from the frequency-doubling 
resonance condition
\begin{equation}
\omega_{s}=2\omega_p \label{freqmatch}
\end{equation}
and makes the phase matching easier to satisfy.

\begin{figure}[t]
%\vspace*{-0.1in}
\centerline{
\input epsf
\setlength{\epsfxsize}{2.5in} \epsffile{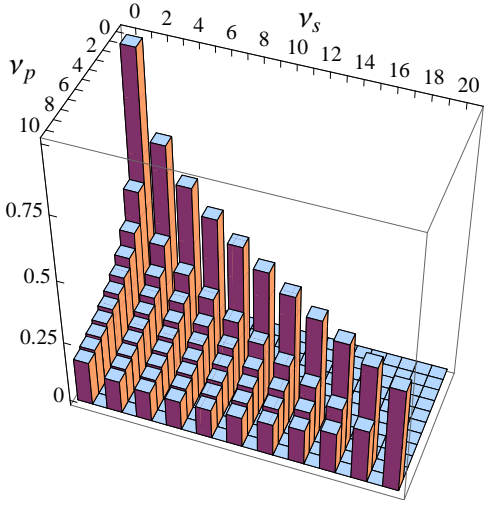} 
}\caption[]{\label{fig:angular}(Color online) Normalized Clebsch-Gordan coefficients
$
\langle m+\nu_p, m+\nu_p ;m, m | 2m+\nu_s,2m\rangle$/$\langle m,m;m, m | 2m,2m\rangle.
$ }
\vspace*{-0.2in}
\end{figure}

The relative SH conversion efficiencies for different WGMs are found from Eq.~(\ref{sigmaL}) using Gaunt's formula \cite{Gaunt29} and treating large factorials in Stirling's approximation.  
Remarkably, for $L$ and $m$ sufficiently large for dispersion equation (\ref{disp}) to remain a good approximation (which is the case in our experiment), the coefficients (\ref{sigmaL}) normalized to $\langle m, m ;m, m | 2m,2m\rangle$ are practically independent of $m$ and therefore of $R$ and $\lambda$. 
This normalization coefficient describes nonlinear coupling of two equatorial modes ($L = m$). The equatorial modes have a single anti-node at $\theta = \pi/2$ and therefore yield the best coupling of the WGMR with the free-space modes, as well as the best overlap with other equatorial modes. To describe non-equatorial modes, we introduce $\nu_p\equiv L_p-m_p\geq 0$ and $\nu_s\equiv L_{s}-m_{s}\geq 0$. The normalized Clebsch-Gordan coefficients are shown in Fig.~\ref{fig:angular}. From this Figure we see that
the second selection rule of (\ref{m}) breaks the symmetry between the SH and degenerate PDC processes. While the former can couple an equatorial pump only to another equatorial mode, the latter can couple the equatorial pump to a variety of modes.

\begin{figure}[t]
%\vspace*{-0.1in}
\centerline{
\input epsf
\setlength{\epsfxsize}{2.3in} \epsffile{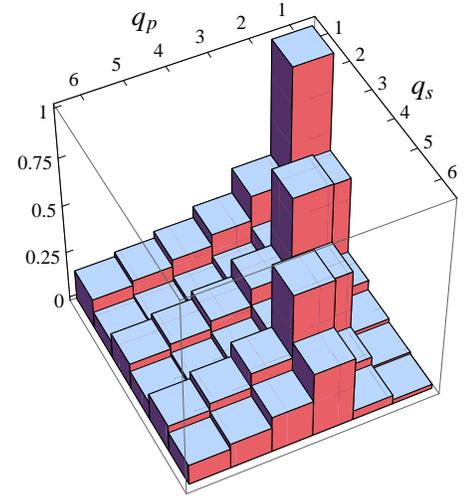} 
}\caption[]{\label{fig:radial}(Color online) Normalized absolute-value coupling coefficients for frequency doubling $\{\omega_p,q_p\}\rightarrow \{2\omega_p,q_s\}$.
}
\vspace*{-0.2in}
\end{figure}

The radial part of integral (\ref{H}) with substituted (\ref{E}) and (\ref{disp}) is evaluated  approximating the large-order Bessel functions by Airy functions. The results are normalized to  the $\{\omega_p,q_p = 1\}\rightarrow \{2\omega_p,q_{s}=1\}$ coupling coefficient, and shown in Fig.~\ref{fig:radial}. These coefficients are practically independent of $L_p$ and $L_{s}$ and therefore of $R$ and $\lambda$. The relative efficiencies of different conversion channels $\{\omega_p,L_p, m_p,q_p \}\rightarrow \{2\omega_p,L_{s}, 2m_p,q_{s}\}$ are found as a product of squares of coefficients from Figs. \ref{fig:angular} and \ref{fig:radial}. 

\begin{figure}[b]
\vspace*{-0.1in}
\centerline{
\input epsf
\setlength{\epsfxsize}{2.3in} \epsffile{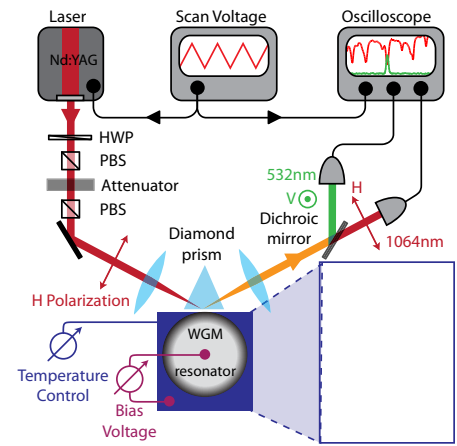} 
}\caption[]{\label{fig:setup}(Color online) Experimental setup. }
\vspace*{-0.2in}
\end{figure}

The multitude of conversion channels allows for a convenient experimental investigation. Our experimental setup is shown in Fig.~\ref{fig:setup}. 
The WGMR is made from a congruent 5\% MgO-doped Lithium Niobate z-cut wafer. Its radius is $R=1.9$ mm and height is approximately 0.5 mm. The rim of the disk is polished into a tapered shape that near the equator can be approximated as a spheroid with radii $R$ and $\rho$. Their ratio $R/\rho\approx 7.5$ is nearly optimal \cite{Strekalov09THz} for coupling of the horizontally polarized pump. 

The pump is a free-space Gaussian beam from a continuous wave Nd:YAG laser (InnoLight Mephisto) with a wavelength of $1064$ nm. It is focused by a microscope objective onto the back side of a diamond prism at an angle that allows for total internal reflection. 
The resonator is placed near the reflected beam's footprint, so that the in- and out-coupling occurs via frustrated total internal reflection. The evanescent field coupling is varied by changing the gap between the prism and the resonator rim. The disk and the prism are mounted on a hot plate whose temperature is stabilized.
The horizontally polarized pump couples to the ordinary polarized modes of the resonator, and the vertically polarized SH is generated into an extraordinary polarized mode. 
Light coupled out of the disk is re-collimated by another objective, and the fundamental and the SH components are separated by a dichroic mirror and focused onto photodetectors. Finally, the signals are monitored by an oscilloscope. 

\begin{figure}[t]
\vspace*{-0.1in}
\centerline{
\input epsf
\setlength{\epsfxsize}{3.4in} \epsffile{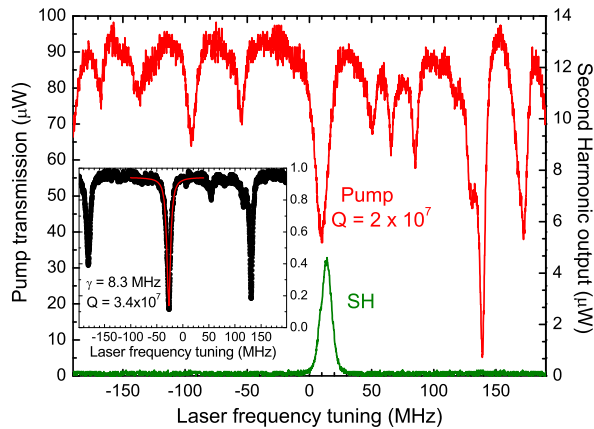} 
}
\vspace*{-0.3in}
\caption[]{\label{fig:wgm}(Color online) Transmitted pump power and the SH signal. The coupling is optimized for high SH output, which corresponds to over-coupled fundamental modes. On the inset: a typical WGM spectrum of a critically coupled resonator at the pump wavelength.}
\vspace*{-0.2in}
\end{figure}

Scanning the laser frequency we observe the pump transmission, which reveals the WGM spectrum at the fundamental wavelength (see Fig.~\ref{fig:wgm}). Fitting the resonances with a Lorentzian function we find a minimal linewidth $\gamma=8.3$ MHz, ($Q=3.4\times 10^7$), and a maximum contrast of 90\%, for a critically coupled resonator. 
 
An electrode attached to the disk allows us to apply a bias voltage to the resonator. The bias shifts the ordinary (pump) and extraordinary (signal) mode frequencies $\omega_p$ and $\omega_{s}$ differently, proportionally to the electro optic coefficients $r_{31}$ and $r_{33}\approx 3.2 r_{31}$ respectively. This differential frequency-tuning of WGMs allows us to achieve condition (\ref{freqmatch}) required for the frequency doubling. 

Changing the resonator temperature has the same effect as changing the bias voltage, since the temperature dependences of the indices of refraction $n_o(T)$ and $n_e(T)$ are different. While the electro-optical tuning is faster, the temperature tuning has a larger dynamic range. 

To find the phase matching temperature for the WGMR, we made a bulk crystal from the same wafer  and determined its phase matching temperature to be $121.7^\circ$C.  This result was consistent with the calculations based on the Sellmeier equation \cite{Schlarb94} and with an earlier experiment \cite{sizmann90}. 
Using the Sellmeier equation together with the spherical WGMR dispersion (\ref{disp}), we calculated the optimal phase matching temperature to be $94.1^\circ$C. 
We assumed the following
parameters: $\omega_p = 2\pi c/1064.3$nm, $m_p \approx 25200$, and $q_{s}=q_p=1$, as this is the most efficient conversion channel
within the accessible temperature and wavelength ranges.

We find the phase matched modes by frequency tuning the extraordinary WGMs at 532 nm relative to the ordinary WGMs at 1064 nm with the bias voltage. We compensate for the frequency shift of the fundamental WGM by the laser central frequency. At some bias voltages the requirement (\ref{freqmatch}) is occasionally satisfied for a pair of modes. Then if the phase-matching conditions (\ref{m}) are also satisfied, SH is generated, see Fig.~\ref{fig:wgm}. For different pump modes this might happen once or several times. However for the majority of the modes it never happens within the explored parameter space. 

\begin{figure}[t]
\vspace*{-0.1in}
\centerline{
\input epsf
\setlength{\epsfxsize}{3.1in} \epsffile{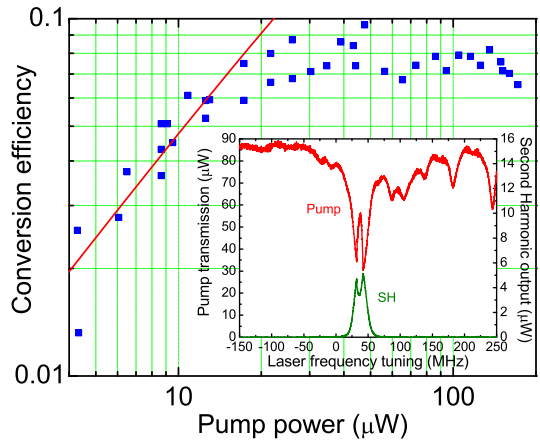} 
}
\vspace*{-0.2in}
\caption[]{\label{fig:efficiency}(Color online) SH conversion efficiency as a function of the in-coupled power: experimental data and theoretical fit \cite{ilchenko04SH} corresponding to the saturation pump power $W_{0} = 3.2$ mW. On the inset: transmitted pump power and the second harmonic signal above the ``self-limiting" threshold which arises at a much lower power than $W_0$.}
\vspace*{-0.2in}
\end{figure}

We observed different conversion efficiency for different WGMs. Fig.~\ref{fig:efficiency} shows the best measured conversion efficiency as a function of the in-coupled pump power. It is worth noting that the {\it detected} SH power is proportional to the power \emph{circulating} in the resonator and to the resonator coupling. The circulating power, on the other hand, is reduced when the coupling increases (due to decreasing Q-factors). As a result, the detected SH power as a function of the coupling has a maximum which corresponds to the over-coupled fundamental WGM.

The theoretical model for the SH generation \cite{AMprivate} in an idealized doubly resonant WGMR has been advanced in \cite{ilchenko04SH}. A key parameter in this model is the saturation power $W_{0}$ which can be considered as an absolute measure of the nonlinear conversion efficiency, independent of the pump power. The experimental value of $W_{0}$ reported in \cite{ilchenko04SH} is 300 mW. In our experiment we measured $W_{0} = 3.2$ mW, see Fig.~\ref{fig:efficiency}. However even this low power greatly exceeds
the theoretical estimate of $W_{0} \approx 6\;\mu$W derived for our resonator based on the theoretical model in ref.~\cite{ilchenko04SH}. This indicates that the spatial overlap between the SH and fundamental WGMs in our experiment deviated from the optimal one by a factor of $\sqrt{W_{0}(theo)/W_{0}(exp)}\approx 0.043$. To explain this discrepancy we point out that
the diagonal coupling coefficients $L_{s} = 2L_p$ in Fig.~\ref{fig:angular} decrease rather slowly with increasing $L-m$. Also, according to Fig.~\ref{fig:radial}, coupling between WGMs with different radial numbers $q$ remains significant even when the difference is large. Therefore we observe multiple instances of phase matching with non-optimal conversion efficiencies and consequently higher saturation power $W_0$,
than for optimal conditions.
It is difficult to uniquely establish which pair of modes yielded the overlap factor inferred from experiment.

Similar non-optimal conversion
efficiencies were measured in the range of $94^\circ$C to $122^\circ$C.  Evidently, the calculations based on the spherical WGMR dispersion (\ref{disp}) have not been sufficiently accurate for our
non-spherical resonator. Considering a very narrow temperature
width of a WGMR phase matching, of the order of $0.001^\circ$C, empirically searching for
the optimal phase matching within a 30-degree temperature range was
cumbersome. A more accurate initial temperature estimate is preferable.

The conversion efficiencies measured both in \cite{ilchenko04SH} and here (see Fig. \ref{fig:efficiency}) reach the maxima and saturate at much lower in-coupled pump powers than $W_0$. We attribute this to a nonlinear ``self-limiting" effect arising when the circulating pump power exceeds a certain threshold, see the inset in Fig.~\ref{fig:efficiency}. This effect is not related to the photorefractive damage (which can also be observed in our resonator, see \cite{savchenkov06}). While the photorefractive damage causes quasi-permanent changes to the WGM spectra, the ``self-limiting" effect shows purely dynamic behavior as a function of the pump power and coupling. This effect may be related to the onset of self-pulsing or similar dynamics \cite{Drummond80Marte94} and will be discussed elsewhere.

In summary, we present the first natural non-critical type-I phase matching for 1064 to 532 nm optical frequency doubling in a high-Q WGMR. This results in a hundred fold lower saturation pump power. Our theoretical study of phase matching in the spherical geometry of WGMRs has predicted the existence of multiple coupled WGM pairs with varying conversion efficiency consistent with our observations. The study also predicted the asymmetry between the SH and PDC selection rules, which may lead to the competition between these processes and explain the observed ``self-limiting" effect.

The best efficiency of 9\% was achieved at 30 $\mu$W in-coupled CW pump power. 
Such a high-efficiency low-power frequency conversion process is very interesting in the context of nonlinear opto-mechanical coupling \cite{optomech}, nonlinear dynamics \cite{Drummond80Marte94} %( especially when optically cleaned [24]) 
and quantum optics \cite{Drummond81}. Experiments with linear cavities showed reduction of amplitude noise below
the quantum limit at the fundamental \cite{pereira88} and SH \cite{sizmann90} frequencies. The theoretical limit for this reduction has been shown \cite{white96} to be determined by the pump noise.
In the low pump power regime, shot noise and sub-shot noise pump light fields are easily accessible, see e.g. \cite{lassen2007}. 
We plan to exploit the low-power operation regime in nonlinear WGMRs to lift the pump noise limitation and to realize a source of non-classical light.

The authors thank Christian Gabriel, and Drs. Andrea Aiello, Andrey Matsko and Anatoliy Savchenkov for discussions. J.U.F. thanks
``IMPRS for optics and imaging" for the stipend.
D.V.S. and M.L. thank the Alexander von Humboldt Foundation for the Fellowship.

\end{document}